# First heat flux estimation in the lower divertor of WEST with embedded thermal measurements

J. Gaspar[a], Y. Corre[a], M. Firdaouss[a], J. L. Gardarein[b], J. Gerardin[a], J. P. Gunn[a], M. Houry[a], G. Laffont[c], T. Loarer[a], M. Missirlian[a], J. Morales[a], P. Moreau[a], C. Pocheau[a], E. Tsitrone[a] and WEST team*

[a]CEA, Institute for Research on Fusion by Magnetic confinement, 13108 Saint-Paul-Lez-Durance, France
[b]Aix Marseille Univ, CNRS, IUSTI, Marseille, France
[c]CEA, LIST, F-91191 Gif-sur-Yvette, France
* See http://west.cea.fr/WESTteam

The present paper deals with the surface heat flux estimation with thermocouples (TC) and fiber Bragg grating (FBG) embedded in the plasma facing components (PFC) of the WEST tokamak. A 2D heat transfer model combined with the conjugate gradient method (CGM) and the adjoint state is used to estimate the plasma heat flux deposited on the PFC. The plasma heat flux is characterized by the time evolution of its amplitude and spatial shape on the target (heat flux decay length $\lambda_q^t$, power spreading in the private flux region $S^t$ and the strike point location $x_0$). As a first step, five ohmic pulses have been investigated with different magnetic configuration and divertor X-point height varying from 44 to 68 mm from the surface. Despite an outboard shift, the relative displacements of the outer strike point as well as the heat flux decay length derived from the TC/FBG systems are consistent with the magnetic equilibrium reconstruction.

## 1. Introduction

The main objective of WEST is to study the behavior of ITER like Plasma Facing Components (PFCs) and to test the resistance and ageing of these components under high heat loads [1,2]. To achieve these objectives, measurement of the distribution and the amplitude of the heat flux on the divertor PFC is necessary. The divertor X-point configuration allows access to a wide range of heat flux shapes on the PFC by changing the X-point height to the target: the magnetic flux surfaces can be compressed or expanded, as soon as the X-point moves away or gets closer to the target respectively [3].

Plasma breakdown, plasma current ramp-up and diverted plasma have been successfully achieved during two WEST experimental campaigns (C2 and C3a). To derive the heat flux from temperature measurements, the plasma must be in steady state, without sweeping, and the plasma current flat top should last for at least a couple of seconds. A series of ohmic pulses meeting these requirements has been performed with X-point height varying from 44 mm up to 68 mm. During ohmic experiments, the power reaching the divertor is too small, so that the PFC surface temperature is below the threshold for IR thermography measurements on tungsten components [4]. Therefore, we propose here to use embedded thermal sensors in order to characterize the heat load pattern in the lower divertor.

The paper is organized as follows. The thermocouple (TC) and fiber Bragg grating (FBG) diagnostics as well as the heat flux estimation methodology are presented in section 2. A detailed description of the heat flux calculation applied to one ohmic pulse is given in section 3. The heat flux decay length on the targets as well as the positions of the maximum heat flux derived from embedded thermal sensors are compared to the magnetic equilibrium reconstruction, as a function of the X-point height, in section 4.

## 2. Diagnostic set-up: TC and FBG systems

### 2.1 WEST lower divertor

The role of the WEST lower divertor target is to sustain the power conducted through the last closed flux surface to the strike points. The WEST divertor consists of 12 independent toroidal sectors of 30° each composed of 38 plasma facing units. During the WEST phase 1 a mix of ITER-like actively cooled components (bulk W) [5] and non-actively cooled W-coated graphite components is used [3]. The W-coated graphite is divided in two separate components, one in the low field side (LFS/outer) and one in the high field side (HFS/inner).

A total of 20 TCs are in the W-coated graphite components in order to study the heat load pattern on the divertor. Here we present the use of the two series of 4 TCs embedded in the outer PFCs. The TCs are 1mm sheathed type N TC embedded at 7.5mm from the surface in a 1.3mm diameter hole and 10mm depth from the PFC side. The TCS are bonded in the hole with graphite adhesive. The graphite adhesive is attempted to sustain 1370°C much higher than 1200°C which is the maximum temperature admissible for both TC type N and W coating. The use of graphite adhesive with thermocouple have been successfully tested in high heat flux facility up to 900°C.

In WEST, 4 optical fiber temperature sensing probes, each of them including 11 regenerated fiber Bragg gratings equally spaced by 12.5 mm, have been specifically designed and installed in the outer PFCs [6]. The FBGs are temperature transducer based on a

_author's email: jonathan.gaspar@univ-amu.fr_

diffraction gratings photowritten by laser into the core of an optical fiber. The FBGs are monitored in reflection by a spectrometer. The spectrometer spectral range is 120 nm between 1500 and 1620 nm with spectral resolution of 8mm. The Bragg wavelengths of the gratings have been spectrally spaced to avoid spectral overlapping during temperature gradient measurement up to 540°C. The 4 FBGs are embedded at 3.5 and 7 mm from the surface, 2 at each depth, in a side groove of 1.2mm height and 4mm depth, in outer PFCs of the divertor only [6]. The regenerated FBG installed have a long-term use above 900°C (>9000h). The advantages of FBG with respect to a single TC are to be immune to electromagnetic interference and to allow the measurement of temperature at different locations on a single fiber. Each sensing line includes 11 spot temperature measurements equally spaced by 12.5 mm in the poloidal direction. The data acquisition rate is 10 and 20 Hz for the FBG and the TC, respectively.

The time response of the TC and FBG have been evaluated about $\tau=250ms$ in a dedicated set-up. In the heat flux calculation we take into account the sensors time-response by convoluting the temperature calculated at the TC location with the sensor step response. Nevertheless the estimation cannot resolve time evolution faster than 250ms, only a time-averaged value can be extracted.

**2.2 Inverse heat conduction problem and heat flux distribution assumptions**

Over the years several techniques have been developed for the heat flux estimation with embedded measurement, some of them with a forward approach [7,8] or with an inverse approach [9,10]. Here we extend the inverse approach described in [10] by using multiple sensors in order to estimate the different parameters which characterize the heat flux spatial distribution.

The inverse heat conduction problem consists in the determination of the surface heat flux ($\phi$) minimizing the discrepancy between the output of a heat conduction problem giving the PFC bulk temperature and the temperature measurements provided by the 4 TCs or the 11 Bragg gratings. The conjugate gradient method combined with the adjoint state is the optimization process used to solve the inverse problem [11,12].

Eich et al. showed for JET and ASDEX-Upgrade divertors [11] that the spatial distribution of the heat flux can be expressed by a heuristic formulation, built with IR thermography during carbon-wall operations, defined by:

$$\phi(x,t) = \phi_M(t) exp\left[\left(\frac{S^t}{2\lambda_q^t}\right)^2 - \frac{x-x_0}{\lambda_q^t}\right] \times erfc\left(\frac{S^t}{2\lambda_q^t} - \frac{x-x_0}{S^t}\right) + \phi_{BG}(t) \quad (1)$$

where $x$ is the target coordinate, $x_0$ is the strike point location, $\lambda_q^t$ is the local heat flux decay length on the target (i.e. not mapped to the outboard midplane), $S^t$ is the heat flux spreading factor on the target, $\phi_M(t)$ the time evolution of the maximal heat flux and $\phi_{BG}(t)$ the background heat flux radiated by the plasma.

Using this expression as a priori in the IHCP we can estimate simultaneously the unknowns $\phi_M(t)$, $\lambda_q^t$, $x_0$ and $\phi_{BG}(t)$ from the TC and FBG measurements. $S^t$ can also be estimated with the FBG measurements thanks to the high number of spot measurements (11 versus 4 for FBG and TC, respectively).

The uncertainties on the estimates using this approach for each diagnostic have been studied in previous work [12,13] with numerical data. These works have shown uncertainties of 8% for $\lambda_q^t$, 10% for $S^t$ and less than 1mm for $x_0$ (but due to the Bragg grating size we will consider an uncertainty of 2mm).

**3. TC, FBG data and estimated heat flux (Ω pulse)**

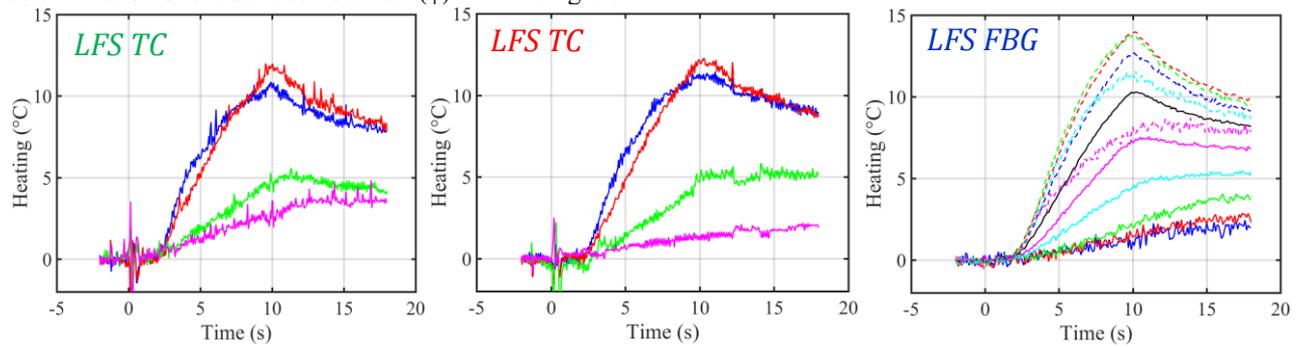

Fig. 1. Temporal evolution of the heating as measured by 2 series of 4 TC and one FBG for the pulse #53237.

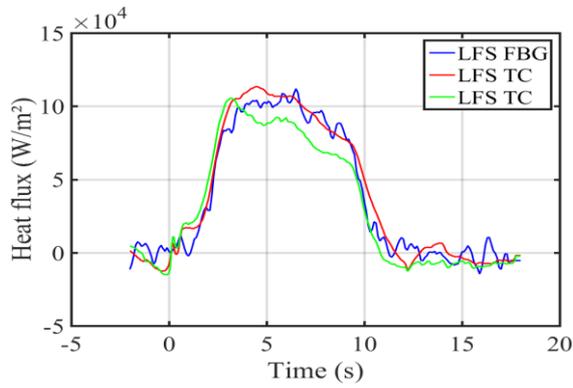

Fig. 2. Time evolution of the estimated heat fluxes with embedded measurements (#53237).

Figure 1 shows the heating of the lower divertor as measured with the different embedded measurements during WEST ohmic plasma with plasma current of Ip = 500kA and toroidal magnetic field Bt = 3.7 T (#53237). From the left to the right, the plots show the two series of the 4 outer TC and one FBG (sensors locations in figure 5). The two diagnostics have good performance in terms of noise with a standard deviation of 0.3 and 0.2°C for the TC and FBG, respectively. The signal noise ratio remains good even for the low heating reported during ohmic experiment. The comparison of the TC and FBG at same poloidal location (blue, red, green and magenta correspond to the FBG N°10, 7, 4 and 1, respectively) shows equivalent heating for the two diagnostics.

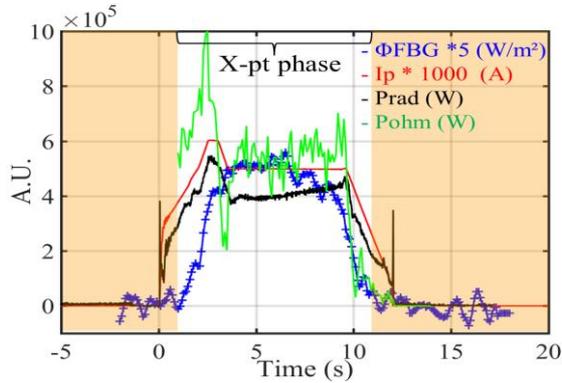

Fig. 3. Estimated heat flux (blue), plasma current ($I_p$), ohmic and radiated power as function of time (#53237).

From these data we can compute the heat flux deposited by the plasma on targets. Figure 2 shows the time evolution of the estimated heat fluxes for the different embedded diagnostics. The time evolution of the heat flux is consistent with the ohmic power and the X-point diverted phase as illustrated in figure 3. The mean amplitude of the estimated heat fluxes during the plasma current flat top phase is found to be ~100 kW/m². The heat flux is small which is consistent with highly radiative ohmic plasma discharges studied here, the radiated power fraction is about 76%.

## 4. Heat flux distribution as function of the magnetic configuration

Table 1 summarizes the plasma parameters of the 5 ohmic pulses studied, the values have been average during the plasma current flat top phase. The pulses have been selected due to the different X point height achieved during these discharges. Based on the magnetic equilibrium reconstruction with the EQUINOX code [10], the variation goes from 44.2mm for the pulse #52665 to 68.5mm for the pulse #53237.

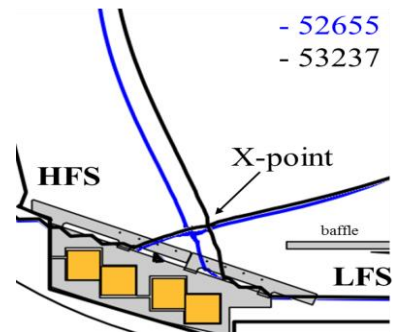

Fig. 4. Magnetic reconstruction (EQUINOX) for the two extreme X-point height.

Figure 4 shows the magnetic reconstructions of EQUINOX near the divertor for the two extreme cases. The increase of the X-point height can be seen as well as the two strike point of the separatrix on the inner and outer PFCs, at the HFS and LFS, respectively.

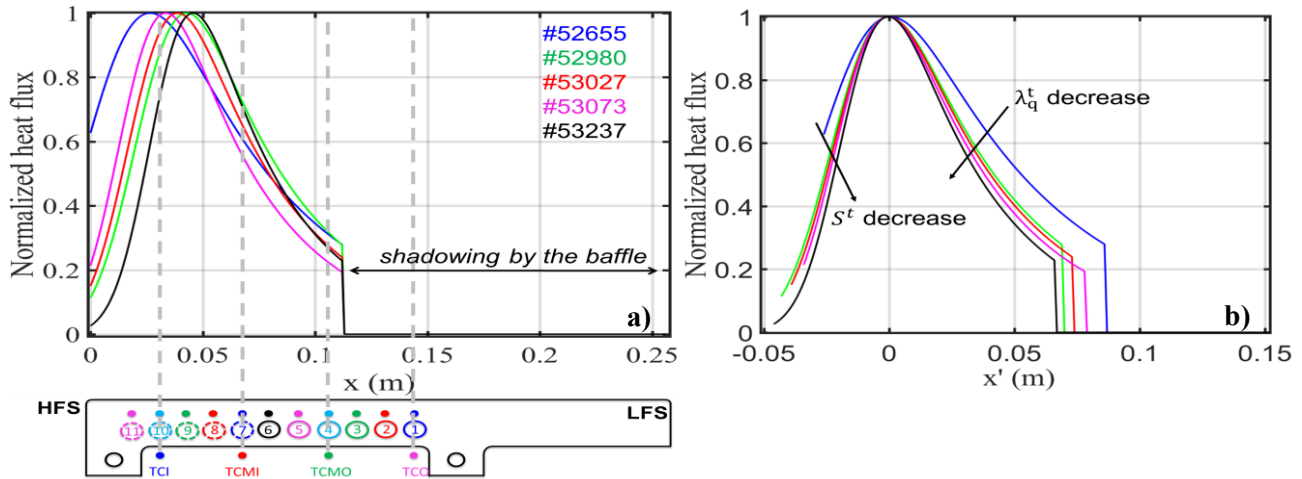

Fig. 5. a) Normalized heat flux as function of space for the 5 investigated ohmic pulses and sensor localization, b) Maximal heat fluxes transposed at the same location.

Figure 5 shows the estimated heat fluxes distribution on PFC for the series of 5 pulses with X-point height varying from 44mm up to 68mm. In order to keep only the heat flux coming from the scrape-off layer, the normalized heat fluxes have been computed by:

$$\bar{\phi}' = \frac{\bar{\phi}-\bar{\phi}_{BG}}{max(\bar{\phi}-\bar{\phi}_{BG})} \text{ mean values during Ip flat top} \quad (1)$$

The change of the heat flux distribution with respect to the magnetic configuration is clearly seen in the figure 5 b). Both parameters, $\lambda_q^t$ and $S^t$, decrease when the X-point height increases and the magnetic flux expansion decreases simultaneously. Figure 6 a) plots the estimated $\lambda_q^t$ and $S^t$ as function of the X-point height. The magnetic expansion, fx, is computed with respect to the outboard mid-plane: it varies from 5.3 down to 3.7 (at the separatrix) when X-point height is 68mm. This confirms that for similar plasma conditions we can compress or expand the heat flux on the target by tuning the X-point height [3].

Figure 6 b) shows the outer strike point location (x₀) versus the X-point height. The values are estimated with the FBG/TC measurements and with the two magnetic reconstruction codes used for WEST (EQUINOX [14] and VACTH [15]). The displacement of the strike point is consistent with the evolution of the X-point height. The SP is moving toward the outer side linearly when the X-point is moved away from the target (see figure 4). However, a mean shift of 23mm towards the baffle (outer side) is observed between EQUINOX and thermal measurements. On the same time, VACTH estimates a lower X-point height than EQUINOX. The discrepancies between the two magnetic reconstruction codes and also with the TC/FBG observations is under investigation. Errors on the passive currents in the structure used in the magnetic reconstructions could be a part of the explanation. Looking at the peak heat flux, the shift is less pronounced (figure 5 a)) due to the simultaneous reduction of $\lambda_q^t$ and $S^t$ (consequence of the equation 1). As a result, the shift of the peak heat flux is about 2 cm on the LFS.

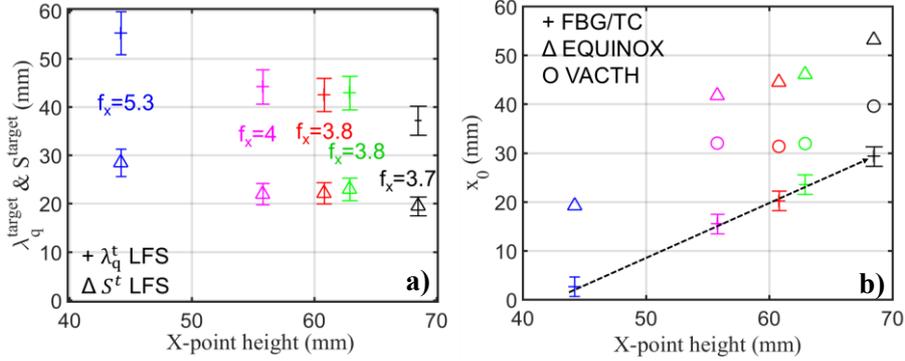

Fig. 6. a) $\lambda_q^t$ and $S^t$ as function of X-point height, b) strike point location (x₀) as function of X-point height for LFS.

| Pulses | Ip (kA) | BT (T) | X-point height (mm) | Pohm (MW) | frad (%) | nl1 ($10^{19}m^{-2}$) divertor | nl3 ($10^{19}m^{-2}$) central | $\phi_{Max}$ LFS (kW/m²) | $\lambda_q^t$ LFS (mm) | $S^t$ LFS (mm) |
|---|---|---|---|---|---|---|---|---|---|---|
| 52655 | 700 | 3.7 | 44.2 | 1 | 86 | 0.1 | 1.6 | 63.5 | 55.3 | 28.5 |
| 53073 | 600 | 3.3 | 55.8 | 0.84 | 77.2 | 0.1 | 2.45 | 76 | 44.2 | 22 |
| 53027 | 600 | 3.6 | 60.8 | 0.79 | 84.2 | 0.125 | 1.2 | 72 | 42.5 | 22.1 |
| 52980 | 594 | 3.7 | 62.9 | 0.825 | 79.1 | 0.125 | 1.25 | 66 | 42.9 | 23 |
| 53237 | 500 | 3.7 | 68.5 | 0.56 | 76.4 | 0.3 | 3 | 100 | 39 | 19.5 |

Table. 1. Plasma parameters and estimated values for the 5 ohmic pulses.

## 4. Conclusion

The WEST embedded PFC diagnostics (TC and FBG) have allowed to estimate heat fluxes on the divertor target for ohmic plasmas with low temperature heating, where no IR measurements are available. The heat load pattern has been investigated in terms of peak heat flux position and broadening as a function of the X-point height. The displacement of the peak heat flux is consistent with the magnetic equilibrium reconstruction. This demonstrates that although the input power is modest, both the TC and the FBG diagnostics are very sensitive to the location of the strike points. This shows that these diagnostics will be very complementary to the IR monitoring system that will be available for surface temperature and heat flux analsyis on the surface of the divertor target for larger input power.

The next experimental campaign will be the opportunity to investigate higher X-point height (up to 90mm) and consolidate these observations with optimized plasma parameters (same plasma current, magnetic field and divertor density) and additional heating power (L-mode experiments).

## Acknowledgments

This work has been carried out thanks to the support of the A*MIDEX project (n°ANR-11-IDEX-0001-02) funded by the "Investissements d'Avenir" French

Government program, managed by the French National Research Agency (ANR).

This work has been carried out within the framework of the EUROfusion Consortium and has received funding from the Euratom research and training programme 2014–2018 under grant agreement No 633053. The views and opinions expressed herein do not necessarily reflect those of the European Commission and ITER Organization